\documentclass[aps,pra,9pt,twocolumn,showpacs,superscriptaddress]{revtex4-1}

\usepackage[utf8]{inputenc}		
\usepackage{amsmath}
\usepackage{latexsym}
\usepackage{amssymb}
\usepackage{bm}
\usepackage{graphics,epstopdf}
\usepackage{color}
\usepackage{hyperref}

\usepackage{newlfont}
\usepackage{amsfonts}
\usepackage{amsthm}
\usepackage{graphicx}
\usepackage{epsfig}
\usepackage{lipsum, babel}

\newcommand{\ket}[1]{|{#1}\rangle}
\newcommand{\bra}[1]{\langle{#1}|}

\newcommand{\ketbra}[2]{|{#1} \rangle \langle {#2} |}

\usepackage{times}

\newcommand{\be}{\begin{equation}}
\newcommand{\ee}{\end{equation}}
\newcommand{\bc}{\begin{center}}
\newcommand{\ec}{\end{center}}
\newcommand{\bea}{\begin{eqnarray}}
\newcommand{\eea}{\end{eqnarray}}
\newcommand{\ba}{\begin{array}}
\newcommand{\ea}{\end{array}}

\newcommand \modu[1]{ \left\lvert#1\right\rvert}

\begin{document}

\title{ Teleportation of quantum coherence}
\author{Sohail $^{1}$, Arun K Pati $^{2}$, Vijeth Aradhya $^{3,4}$, Indranil Chakrabarty $^{2,3}$, Subhasree Patro $^{3,5,6}$}
\affiliation{$^{1}$Quantum Information and Computation Group, Harish-Chandra Research Institute,  A CI of Homi Bhabha National
Institute, Chhatnag Road, Jhunsi, Prayagraj - 211019, India.\\
$^{2}$ Centre for Quantum Science and Technology,\\International Institute of Information Technology-Hyderabad, Gachibowli, Hyderabad, India.\\ 
$^{3}$Center for Security, Theory and Algorithmic Research\\International Institute of Information Technology-Hyderabad, Gachibowli, Hyderabad, India.\\
$^{4}$ Department of Computer Science, National University of Singapore - 117417 Singapore.\\
$^{5}$ Utrecht University, 3584 CS Utrecht, The Netherlands\\
$^{6}$ QuSoft, Centrum Wiskunde $\&$ Informatica, 1098 XG Amsterdam, The
Netherlands.}

\begin{abstract}
We investigate whether it is possible to teleport the coherence of an unknown quantum state from Alice to Bob by communicating a lesser number of classical bits in comparison to what is required for teleporting an unknown quantum state. We find that we cannot achieve perfect teleportation of coherence with one bit of classical communication for an arbitrary qubit. However, we find that if the qubit is partially known, i.e., chosen from the equatorial and polar circles of the Bloch sphere, then teleportation of coherence is possible with the transfer of one cbit of information when we have maximally entangled states as a shared resource. In the case of the resource being a non-maximally entangled state, we can teleport the coherence with a certain probability of success. In a general teleportation protocol for coherence, we derive a compact formula for the final state at Bob's lab in terms of the composition of the completely positive maps corresponding to the shared resource state and joint POVM performed by Alice on her qubit and the unknown state. Using this formula, we show that teleportation of the coherence of a partially known state with real matrix elements is possible perfectly with the help of a maximally entangled state as a resource. Furthermore, we explore the teleportation of coherence with the Werner states and show that even when the Werner states become separable, the amount of teleported coherence is non-zero, implying the possibility of teleportation of coherence without entanglement.  
\end{abstract}
\maketitle
\section{\label{sec:Intro}Introduction}
\noindent Quantum Coherence \cite{Baum} and entanglement \cite{Horo} are two central features in quantum theory that make the theory counterintuitive and can also be utilised as a resource to perform important information theoretic and computational tasks. Motivated by this increasing importance, 
a general study from the perspective of resource theory is being formulated. Several measures based on resource theoretic approaches are available to quantify the coherence and the entanglement present in a quantum system \cite{Baum, Horo, Wint, Stret1, Stret2, Chit, Theu, Mukho, Mitch, Bu, Strlet3, Asboth, Xi, Strlet4, Qi, Chin, Zhu, Zhu1, Korz}. Quantum coherence can be interpreted in several ways, like as a measure of non-classicality in physical systems, a measure of superposition, and as a quantity capturing the wave aspect of the state vector \cite{WaveParticleDuality1, WaveParticleDuality2, WaveParticleDuality3}. It is also observed that perfect cloning and broadcasting of quantum coherence are not possible \cite{We,udit}. Coherence acts as a resource in tasks like quantum algorithms \cite{Anand, Li, Cas}, biological processes \cite{band, wilde} quantum metrology \cite{Metro1,Metro2,Metro3,Metro4,Metro5}, reference frame alignment \cite{Bartl}, thermodynamic tasks \cite{Thermo1,Thermo2,Thermo3,Thermo4,Thermo5}. From the point of view of resource theory, coherence is classified into two classes, namely speakable coherence and unspeakable coherence\cite{Iman}. The quantification of coherence involves a chosen basis. In situations where the labelling of the chosen basis is not important, the relevant notion of coherence is speakable one, and in situations where the labelling or the identity of the basis elements matters, the relevant notion of coherence is unspeakable one. Coherence involved in quantum metrology, reference frame alignment, and thermodynamic tasks are examples of unspeakable coherence, whereas the coherence involved in computational, cryptographic, and communication tasks are examples of speakable coherence. 

\noindent 
In the last two decades, several developments have been made in quantum based technologies. These include quantum repeaters for communicating over large distances \cite{rep}, quantum  teleportation \cite{Bennett}, broadcasting of entanglement \cite{Bro1,Bro2, Bro3, Bro4}. These were introduced in the process with the vision of having quantum networks \cite{Steph,Jacob, Per}.  In particular, "Quantum teleportation" is one of the most fascinating discoveries of the 20th century \cite{Bennett}.  The entangled states which are useful for teleportation are identified as well as detected in this process \cite{I1, I2, I3, D1}. Research was also done to implement the teleportation process, if not perfectly but at least with a certain probability of success \cite{Agrawal}. 
Recently, it has also been utilized to understand phenomena like closed time-like curves \cite{Kumar}.  
It has been demonstrated in the laboratory with the help of various resources like photonic qubits, nuclear magnetic resonance (NMR)\cite{28}, optical
modes \cite{29,30,31,32,33,34,35,36,37} , atomic ensembles \cite{38,39,40,41}, trapped atoms \cite{42,43,44,45,46}, and
solid state systems\cite{47,48,49,50}. Efficient teleportation was also
achieved in terms of distance \cite{21,22}.

\noindent In this article, in general, we address the question of whether we can teleport the coherence of a quantum state, with a lesser number of classical bits than what is required to teleport the state itself. It is already known that teleportation of an unknown state is possible with the transfer of two classical bits of information. The process of teleportation of a known state is known as remote state preparation, and it requires the transfer of one bit of information for the process to be (only for equatorial qubit) perfect \cite{RSP}. It is obvious that once the entire information of the state is recreated in a different location, the coherence of the state is also transferred. There are many instances where it has been seen that the coherence of the state is not the entire information of the state \cite{We}. So it becomes a natural question: will the transfer of coherence in either known or unknown cases require a lesser number of bits? We find  in this article that we are able to teleport the coherence of a partially known state with one bit of classical communication and with a maximally entangled resource when the qubits are taken from equatorial and polar circles. In addition to that, we show that if we start with a non-maximally entangled state as a resource, then we can teleport coherence probabilistically. Here we came up with a compact formula for the final state at Bob's lab. This is done in terms of the composition of the completely positive maps corresponding to the shared resource state and the joint POVM performed by Alice to her qubit and the unknown particle. With this, we also find out the amount of coherence that is teleported when we have the shared resource state as the maximally entangled mixed states and Werner states. We also discuss the general teleportation of coherence for an initial mixed state as an input state at Alice's lab. The results of the article stand out from the perspective of transferring the coherence of a coherent state to a state where there is no coherence. This is important in a quantum network since coherence is a useful resource. In a quantum network, one node can teleport the coherence to another node where it is required. Here we show that this is possible without actually transferring the entire information of the state. Interestingly,
we are able to show this can be done much more cost-effectively with a
lesser number of classical bits if the states are from the equatorial or polar
circles. Our results are useful from the perspective of work extraction as well. Korzekwa et al. \cite{Korzekwa} have shown the existence of thermal machines that can extract work from coherence arbitrarily well. These machines only have to operate on individual copies of a state. When Alice and Bob share a maximally entangled state, before teleportation of coherence, the coherence at Bob's side is zero, and hence no work extraction is possible. However, after the teleportation, Bob can use the teleported coherence for work extraction.  

\noindent The paper is organized as follows: In Section II, we study the possibility of the teleportation of quantum coherence of a pure state with maximally as well as non-maximally entangled resources using
fewer cbits. In Section III, we investigate the same for an arbitrary mixed state with maximally entangled mixed state and Werner state as resources and demonstrate the possibility of teleporting some fraction of the initial coherence without entanglement. We conclude the article in section IV.
\\

\section{\label{sec:CohTele}Teleportation of Quantum Coherence With Pure Entangled States as a Resource}
\noindent In this section, our aim is to address the question of whether we can perfectly teleport the coherence of an unknown state with a lesser number of classical bits than what is required for the teleportation of the state. Before we present our protocol, let us briefly recapitulate the notion of quantum coherence.

Let ${\mathcal{H}} =\mathbb{C}^d$ be the $d$ dimensional Hilbert space associated with the qudit of our consideration. Here, $S(\mathcal{H})$ is the set of positive trace-class linear operators, with trace being $1$  on the Hilbert space $\mathcal{H}$. For a state $\rho \in S(\mathcal{H} )$, the coherence of the state is given by  $C_{l_{1}}(\rho) = \sum^{d}_{i\neq j} |\langle i| \rho | j \rangle|$, where $\{|i\rangle\}$ is an orthonormal basis of $\mathcal{H}$ . It is clear from the definition that the $l_1$-norm of coherence is basis-dependent. The states which are diagonal in the matrix representation with respect to this chosen basis are defined as incoherent states. Therefore, the incoherent states are of the form $\rho=\sum_{i=1}^d \lambda_i \ket{i} \bra{i}$ .
Let us denote the set of all incoherent states by ${I} \subset S(\cal H)$. Quantum operations on a physical system are mathematically represented by a completely positive and trace-preserving (CPTP) linear map $\Phi:B(\cal H) \rightarrow B(\cal H)$ with $B(\cal H)$ being the set of bounded linear operators on $\mathcal{H}$. It is well known that the action of a CPTP map $\phi$ on a quantum state $\rho$ can be represented as $\Phi(\rho)= \sum_{i}^n K_i \rho K_i^\dagger $, where $\sum_{i}^n   K_i^\dagger K_i = \mathbb{I}$ with $\mathbb{I}$ being the identity operator on $\mathcal{H}$. This representation of a CPTP map is known as the operator-sum representation, and the linear operators $K_i: \cal H \rightarrow \cal H$ are called Kraus operators. Given a CPTP map, its Kraus representation is not unique. A quantum operation $\Phi$ is said to be incoherent if $K_i I K_i ^\dagger \subset I$ for all $i=1,..,n$.
\\
Now any measure of coherence $C$ is defined to be a functional from the set of quantum states to the non-negative real numbers such that the following constraints are obeyed.\\
i) $C(\rho)=0$ for all $\rho \in I$, i.e., for all incoherent states, the measure of coherence is zero.\\
ii) The measure of coherence should not increase under incoherent operations. If $\Lambda$ is an incoherent operation, then $C(\Lambda(\rho)) < C(\rho) $ \\
iii) It is strongly monotonic: It does not increase under selective incoherent operations on average, i.e., $C(\sum_i p_i \rho_i) < C(\rho)$ for all $\{ K_i \}$ with  $\sum_{i}   K_i^\dagger K_i = \mathbb{I}$ and $K_i I K_i ^\dagger \subset I$, where $p_i= Tr(K_i \rho K_i ^\dagger)$ and $\rho_i =\frac{K_i \rho K_i ^\dagger}{p_i} $.
\\


\noindent  Let us denote the system whose state is the unknown quantum state $\ket{\psi}$ as ''system 1", one part of the shared state at Alice's lab as ''system 2" and the other part at Bob's side as "system 3". We imagine that Alice and Bob share an entangled pair in the maximally entangled state $\ket{\Phi^+}_{23}=\frac{1}{\sqrt{2}}(\ket{00}+\ket{11})$. An unknown state $\rho_1 =  |\psi \rangle \langle \psi| $ is given to Alice, and it has $C(\rho_1)$ amount of coherence.
We consider a general scenario where Alice performs a joint measurement $\Pi_i$ which is a positive operator-valued measure (POVM) on the input particle and half of the entangled pair, and $\sum_i \Pi_i =\mathbb{I}$. Now, depending on the measurement outcomes, the reduced state of Bob, which is unnormalised, can be written as
\begin{eqnarray}
    \rho_3^{(i)} = Tr_{12} [ \Pi^{(i)}_{12} (\rho_1 \otimes  |\Phi \rangle \langle \Phi|) \Pi^{(i)}_{12}].
\end{eqnarray}
Where we have adopted the notation that $\Pi^{(i)}_{12}$ means $\Pi^{(i)}_{12}\otimes \mathbb{I}$.
\\
\noindent After Alice communicates the measurement outcome to Bob via a classical channel, Bob performs a local unitary, and we expect the coherence of the state at Bob's location to have the same coherence as that of the input state, i.e., $C(\rho_3^{(i)}) = C(\rho)$. If this condition holds, then we say that teleportation is complete. \\
\noindent In this article, we start with a two-level system as an input and an arbitrary two-qubit shared state as a resource. In particular, we also consider both maximally and non-maximally entangled states as shared resources. For an unknown qubit on Alice's side, we ask the question whether we can teleport the coherence of the qubit to Bob's side with 1 bit of classical communication or not. In each of these cases, we find that we cannot do it perfectly with one bit of classical communication universally. However, if we have a maximally entangled state as a resource and we have partial knowledge about the state of the input qubit, i.e., it is chosen from one of the two great circles (equatorial and the polar circle) on the Bloch sphere, coherence teleportation is possible perfectly with the use of one cbit. \\ 

\subsection{Teleportation of Coherence of a Qubit with Maximally entangled state as a resource} 

The main idea is to teleport the coherence of an unknown state $\ket{\psi}$ from Alice's lab to Bob's lab, rather than the state itself. If, with the help of only one cbit, Alice can teleport a state at Bob's lab that has the same $l_1$-norm coherence as the unknown state, $|\psi\rangle = \alpha|0\rangle + \beta|1\rangle \in \mathbb{C}^2$, where $\alpha$ and $\beta$ are complex numbers such that $|\alpha|^2+|\beta|^2 = 1$, then the purpose is fulfilled.\\
The unknown quantum state $\ket{\psi}$ can be represented as
\begin{eqnarray}
 |\psi\rangle=\cos(\theta/2)|0\rangle+\sin(\theta/2)e^{i\phi}|1\rangle
\end{eqnarray}
where $\theta$ and $\phi$ with $0 \leq \theta \leq \pi$ and $0 \leq \phi < 2\pi$ (the exact value of $\theta$ and $\phi$ are not known to us). The coherence of the state $\ket{\psi}$ in terms of these parameters with respect to the computational basis $\{\ket{0},\ket{1}\}$ is
$C(\psi)=2|\alpha\beta|= \sin \theta$.
Our aim is to have $C(\psi)= sin \theta$ amount of coherence at Bob's lab at the end of the protocol.

\noindent In the standard teleportation protocol,  Alice and Bob share a maximally entangled state 
$|\Phi \rangle_{23} =|\Phi^+ \rangle_{23} =\frac{1}{\sqrt[]{2}}(|0 0 \rangle_{23}+|11\rangle_{23})$ as resource. It is evident that if we teleport the unknown state itself, then the coherence also gets teleported. The $l_{1}$-norm coherence of all those states (before Bob's unitary operation) is $2\cos(\theta/2)\sin(\theta/2)$ which is the same as the coherence of the initial state. However, this still requires two bits of classical information.
  
\noindent However, the interesting question will be  whether perfect teleportation of coherence can be done with lesser number of cbits. The main idea is that instead of doing Bell measurement, we do POVM measurement by adding different Bell projectors in various combinations. Given four Bell projectors, there are only three different ways we can add them to form a complete measurement. Let us consider the following situation where the state shared between Alice and Bob is
\begin{eqnarray}
 \ket{\Phi^+}=\frac{1}{\sqrt{2}}(\ket{00}+\ket{11}),
\end{eqnarray}


\noindent \textbf{Case I:}
Let Alice and Bob share a maximally entangled state $\ket{\Phi^+}$. The combined state of the input and the shared state can be expressed as 
\begin{eqnarray}
 \ket{\psi} \otimes \ket{\Phi^+} = \frac{1}{2}\sum_{i=0}^{3} \ket{B_i} \otimes u_i \ket{\psi},
\end{eqnarray}
where $\ket{B_i}$ are the four mutually orthogonal Bell states and $u_i$ are the local unitary operations ($I, \sigma_x, \sigma_y, \sigma_z)$.

Alice instead of performing Bell measurement, performs the following POVM on the input and half of the entangled pair 
 \begin{eqnarray}
  \Pi^{(0)}_{12} =\ket{\Phi^+}\bra{\Phi^+} + \ket{\Psi^+}\bra{\Psi^+},\label{povmI}
 \end{eqnarray}
 \begin{eqnarray}
    \Pi^{(1)}_{12}=\ket{\Phi^-}\bra{\Phi^-}+\ket{\Psi^-}\bra{\Psi^-}.\label{povmII}
 \end{eqnarray} 
 where $\ket{\Phi^\pm}$ and $\ket{\Psi^\pm}$ are standard Bell states given by
\begin{eqnarray}
&& \ket{\Phi^-}=\frac{1}{\sqrt{2}}(\ket{00}-\ket{11}),\\
&& \ket{\Psi^+}=\frac{1}{\sqrt{2}}(\ket{01}+\ket{10}),\\
&& \ket{\Psi^-}=\frac{1}{\sqrt{2}}(\ket{01}-\ket{10}).
\end{eqnarray}

After Alice performs the measurement given by the POVM element $\Pi^{(0)}_{12}$ she communicates her result to Bob. The state of the particle at Bob's lab is given by
\begin{eqnarray}
\nonumber
\rho_3^{(0)} = \frac{1}{p_0} Tr_{12} [ \Pi^{(0)}_{12} ( \ket{\psi} \bra{\psi} \otimes \ket{\Phi^+}\bra{\Phi^+}  ) \Pi^{(0)}_{12} ],
\end{eqnarray}
where $p_0 = Tr_{123} [ \Pi^{(0)}_{12} ( \ket{\psi} \bra{\psi} \otimes \ket{\Phi^+}\bra{\Phi^+}  ) \Pi^{(0)}_{12} ]$.
Therefore, Bob's state can be expressed as 
\begin{eqnarray}
\nonumber
 \rho_3^{(0)}=\frac{I}{2} + Re(\alpha \beta^*) (\ket{0}\bra{1}+ \ket{1}\bra{0} ).
\end{eqnarray}

For the measurement with the POVM element $\Pi^{(1)}_{12}$, the state of the particle at Bob's lab is given by
\begin{eqnarray}
\nonumber
\rho_3^{(1)} = \frac{1}{p_1} Tr_{12} [ \Pi^{(1)}_{12} ( \ket{\psi} \bra{\psi} \otimes \ket{\Phi^+}\bra{\Phi^+}  ) \Pi^{(1)}_{12} ],
\end{eqnarray}
where $p_1 = Tr_{123} [ \Pi^{(1)}_{12} ( \ket{\psi} \bra{\psi} \otimes \ket{\Phi^+}\bra{\Phi^+}  ) \Pi^{(1)}_{12} ] $.
Therefore, Bob's state can be expressed as 
\begin{eqnarray}
\nonumber
 && \rho_3^{(1)}= \frac{I}{2} -Re(\alpha \beta^*)(\ket{0}\bra{1}+\ket{1}\bra{0}).
\end{eqnarray}
 The coherence of the state at Bob's site for both the case is
 \begin{eqnarray}
 \nonumber
   && C_{l1}=2\lvert Re(\alpha \beta^*) \rvert\\
   \nonumber
     &&=cos(\phi) sin(\theta)\\
     \nonumber
    && \neq sin(\theta).
 \end{eqnarray}
 Interestingly, this equality will hold when $\cos(\phi)=1$. This condition actually tells us that if we have partial information about the state of the qubit, i.e., it comes from the equatorial circle, then the teleportation of the coherence will be possible with the help of only one cbit.\\
 \\
\textbf{Case II:}
In this case, Alice performs the following POVM on her part,
 \begin{eqnarray}
 \nonumber
  E^{(0)}_{12}=\ket{\Phi^+}\bra{\Phi^+}+\ket{\Psi^-}\bra{\Psi^-}
 \end{eqnarray}
 \begin{eqnarray}
 \nonumber
  E^{(1)}_{12}=\ket{\Phi^-}\bra{\Phi^-}+\ket{\Psi^+}\bra{\Psi^+}
 \end{eqnarray}  
The state at Bob's part after Alice performs the measurement with POVM element $E^{(0)}_{12}$ and communicates to Bob, is given by
\begin{eqnarray}
\nonumber
\rho_3^{(0)} = \frac{1}{p_0} Tr_{12} [ E^{(0)}_{12} ( \ket{\psi} \bra{\psi} \otimes \ket{\Phi^+}\bra{\Phi^+}  ) E^{(0)}_{12} ],
\end{eqnarray}
where $p_0 = Tr_{123} [ E^{(0)}_{12} ( \ket{\psi} \bra{\psi} \otimes \ket{\Phi^+}\bra{\Phi^+}  ) E^{(0)}_{12} ] $.
Therefore, Bob's state can be expressed as 
\begin{eqnarray}
\nonumber
  && \rho^{(0)}_3=\frac{I}{2}+i Im(\alpha \beta^*)(\ket{0}\bra{1}-\ket{1}\bra{0}).
\end{eqnarray}
\\
The state at Bob's part after Alice performs the measurement with POVM element $E^{(1)}_{12}$ and communicates to Bob, is given by
\begin{eqnarray}
\nonumber
\rho_3^{(1)} = \frac{1}{p_1} Tr_{12} [ E^{(1)}_{12} ( \ket{\psi} \bra{\psi} \otimes \ket{\Phi^+}\bra{\Phi^+}  ) E^{(1)}_{12} ],
\end{eqnarray}
where $p_1 = Tr_{123} [ E^{(1)}_{12} ( \ket{\psi} \bra{\psi} \otimes \ket{\Phi^+}\bra{\Phi^+}  ) E^{(1)}_{12} ] $.
Bob's state in this case can be expressed as 
\begin{eqnarray}
\nonumber
  && \rho^{(1)}_3=\frac{I}{2} -i Im(\alpha \beta^*)(\ket{0}\bra{1}-\ket{1}\bra{0}).
\end{eqnarray}
Now the $l1$-norm coherence of these two states turns out to be:
\begin{eqnarray}
\nonumber
  && C_{l1}(\rho^{(0)}_3) = C_{l1}(\rho^{(1)}_3) = 2|Im(\alpha\beta^*)|{}\nonumber\\
  \nonumber
 &&= 2\sin\phi\cos(\theta/2)\sin(\theta/2)\\
 \nonumber
 && \neq 2\cos(\theta/2)\sin(\theta/2).
 \end{eqnarray}
Interestingly, the equality will hold when $\sin(\phi)=1$. Like the previous case, if the qubit is partially known, i.e., it comes from the polar circle, then the teleportation of coherence is perfectly possible with the use of only one cbit.
\\
\noindent \textbf{Case III:} If Alice performs the following POVM on her part,
\begin{eqnarray}
\nonumber
  F^{(0)}_{12}=\ket{\Phi^+}\bra{\Phi^+}+\ket{\Phi^-}\bra{\Phi^-},
\end{eqnarray}
 \begin{eqnarray}
 \nonumber
  F^{(1)}_{12}=\ket{\Psi^+}\bra{\Psi^+}+\ket{\Psi^-}\bra{\Psi^-}.
 \end{eqnarray}   
 The state at Bob's part after Alice performs the measurement with POVM element $F^{(0)}_{12}$ and communicates to Bob is given by
\begin{eqnarray}
\nonumber
\rho_3^{(0)} = \frac{1}{p_0} Tr_{12} [ F^{(0)}_{12} ( \ket{\psi} \bra{\psi} \otimes \ket{\Phi^+}\bra{\Phi^+}  ) F^{(0)}_{12} ],
\end{eqnarray}
where $p_0 = Tr_{123} [ F^{(0)}_{12} ( \ket{\psi} \bra{\psi} \otimes \ket{\Phi^+}\bra{\Phi^+}  ) F^{(0)}_{12} ] $.
Therefore, Bob's state can be expressed as
 \begin{eqnarray}
 \nonumber
  \rho^{(0)}_3=\frac{I}{2}
 \end{eqnarray}
 \\
  The state at Bob's part after Alice performs the measurement with POVM element $F^{(1)}_{12}$ and communicates to Bob is given by
\begin{eqnarray}
\nonumber
\rho_3^{(1)} = \frac{1}{p_1} Tr_{12} [ F^{(1)}_{12} ( \ket{\psi} \bra{\psi} \otimes \ket{\Phi^+}\bra{\Phi^+}  ) F^{(1)}_{12} ],
\end{eqnarray}
where $p_1 = Tr_{123} [ F^{(1)}_{12} ( \ket{\psi} \bra{\psi} \otimes \ket{\Phi^+}\bra{\Phi^+}  ) F^{(1)}_{12} ] $.
Therefore, Bob's state can be expressed as
\begin{equation*}
    \rho^{(1)}_3=\frac{I}{2} 
\end{equation*}
So, in this case, with the help of one cbit, it is not possible to teleport the coherence of the qubit, even if it is partially known, as the final state has zero coherence. 
\noindent We show that perfect teleportation of coherence is \textit{not} possible universally with maximally entangled state as a resource and POVM measurement on Alice's side with the transfer of one cbit of information. However, if the input qubit happens to be from the equatorial and polar circles, then we can perfectly teleport the coherence of the qubit with the help of one cbit of information. 
\par When the partially known qubit is not from the equatorial and polar circles of the Bloch sphere, it is interesting to see whether Bob can increase the teleported coherence by applying any general unitary transformation on his part after receiving the cbit from Alice so that the coherence of his qubit becomes the same as the original coherence at Alice's side. Consider a general $SU(2)$ matrix,
\begin{equation}
\nonumber
U_{2 \times 2} = \begin{bmatrix}
	a & b \\
  	-b^* & a^*
	\end{bmatrix},
\end{equation}
where $a$ and $b$ are complex numbers such that $|a|^2 + |b|^2 = 1$.\\
\noindent In \textbf{Case 1}, after the application of $U$, the new state denoted by $\rho_{U}$ will be,
\begin{align*}
\begin{split}
\rho_{U} &= U\rho_3^{\Pi_0} U^\dagger\\
         &= U \begin{bmatrix}
	1/2 & Re(\alpha\beta^*) \\
  	Re(\alpha\beta^*) & 1/2
	\end{bmatrix} U^\dagger.
\end{split}
\end{align*}
The $l_1$-norm coherence comes out to be: 
\begin{eqnarray}
\nonumber
 && C_{l1}(\rho_U) = 2\modu{Re(\alpha\beta^*)}\lvert(a^2-b^2)\rvert\\
 &&=2\modu{Re(\alpha\beta^*)}\sqrt{1-(2Re(ab^*))^2} \label{coherence}\\
 \nonumber
 &&\neq 2|\alpha||\beta|.
\end{eqnarray}

Note that the maximum possible value of the quantity $\sqrt{1-(2Re(ab^*))^2}$ is 1. The RHS of Eqn.~(\ref{coherence}) can at most be $ 2Re(\alpha\beta^*)$ which is not equal to $ 2|\alpha||\beta|$,  So the application of unitary operators on Bob's part does not give any significant advantage. Similarly, for the other cases, it can be shown that there is no \textit{general} unitary operator that can take all possible Bob's states to the state which will have the coherence of the original states. Hence, although perfect teleportation of quantum coherence is possible for specific classes of states (states from the equatorial and polar circles), universally it is not possible even if Bob applies any unitary transform on his part.

\subsection{Teleportation of coherence with the help of non-maximally entangled state:}
Here we investigate whether we can teleport the coherence of an unknown state using a non-maximally entangled state and by communicating one cbit of information. We will see that indeed it is possible with a certain probability of success if we have partial knowledge about the state.\\

\noindent Let the state shared between Alice and Bob be the following:
\begin{eqnarray}
  \ket{\Phi}_{AB} = \frac{1}{\sqrt{1+\lvert n \rvert ^2}}(\ket{00}+ n \ket{11}),
\end{eqnarray}
where $n$ is a complex number.
Now let us consider the following four mutually orthonormal vectors, which form a basis of $\mathbb{C}^2 \otimes \mathbb{C}^2$ .
\\
\begin{eqnarray}
\nonumber
 && \ket{\Phi_n^+} = \frac{1}{\sqrt{1+\lvert n \rvert ^2}}(\ket{00}+ n \ket{11}),\\
 \nonumber
 && \ket{\Phi_n^-} = \frac{1}{\sqrt{1+\lvert n \rvert ^2}}(n^*\ket{00}- \ket{11}),\\
 \nonumber
 && \ket{\Psi_n^+} = \frac{1}{\sqrt{1+\lvert n \rvert ^2}}(\ket{01}+ n^* \ket{10}),\\
 \nonumber
 && \ket{\Psi_n^-} = \frac{1}{\sqrt{1+\lvert n \rvert ^2}}(n \ket{01}- \ket{10}).
\end{eqnarray}

Notice that these are non maximally entangled vectors in $\mathbb{C}^2 \otimes \mathbb{C}^2$.\\
 
\noindent \textbf{Case I:} Alice will perform the following POVM to her part:
 \begin{eqnarray}
 \nonumber
   && e^{(0)}_{12}=\ket{\Phi_n^+}\bra{\Phi_n^+}+\ket{\Psi_n^-}\bra{\Psi_n^-},\\
   \nonumber
   && e^{(1)}_{12}=\ket{\Phi_n^-}\bra{\Phi_n^-}+\ket{\Psi_n^+}\bra{\Psi_n^+}.
 \end{eqnarray}
The state at Bob's part after Alice performs the measurement with POVM element $e^{(0)}_{12}$ and communicates to Bob, is given by
\begin{eqnarray}
\nonumber
\rho_3^{(0)} = \frac{1}{p_0} Tr_{12} [ e^{(0)}_{12} ( \ket{\psi} \bra{\psi} \otimes \ket{\Phi_n^+}\bra{\Phi_n^+}  ) e^{(0)}_{12} ],
\end{eqnarray}
where $p_0 = Tr_{123} [ e^{(0)}_{12} ( \ket{\psi} \bra{\psi} \otimes \ket{\Phi_n^+}\bra{\Phi_n^+}  ) e^{(0)}_{12} ] $.
Therefore, Bob's state can be expressed as
\begin{eqnarray}
\nonumber
  \rho_3^{(0)}  && =\frac{1}{1+\lvert n \rvert^4} \ket{0}\bra{0}+\frac{\lvert n \rvert^4}{1+\lvert n \rvert^4} \ket{1}\bra{1}+\\
  \nonumber
    &&\frac{\lvert n \rvert^2}{1+\lvert n \rvert^4} 2i Im(\alpha \beta^*) \ket{0}\bra{1}-\\
    \nonumber
    && \frac{\lvert n \rvert^2}{1+\lvert n \rvert^4} 2i Im(\alpha \beta^*) \ket{1}\bra{0}.
\end{eqnarray}

\noindent The probability of success for the POVM $e^{(0)}_{12}$ to be clicked is given by $p_{0}=\frac{1+\lvert n \rvert^4}{(1+\lvert n \rvert^2)^2}$.\\
 \\
\noindent The coherence of the state at Bob's side is given by
 \begin{eqnarray}
 \nonumber
   C_{l_1}(\rho_3^{(0)})&&=\frac{4 \lvert n \rvert^2}{1+\lvert n \rvert^4} \lvert Im(\alpha \beta ^*) \rvert\\
   \nonumber
   &&=\frac{2 \lvert n \rvert^2}{1+\lvert n \rvert^4} 2\sin\phi\cos(\theta/2)\sin(\theta/2)\\
   \nonumber
 && \neq 2\cos(\theta/2)\sin(\theta/2).
 \end{eqnarray}
 Unlike the case discussed in the previous section, even if the state of the input qubit is partially known, i.e., it is from the polar circle of the Bloch sphere, the coherence is not perfectly teleported. It is $\frac{2 \lvert n \rvert^2}{1+\lvert n \rvert^4}$ times the original coherence.\\
 
 \noindent The state at Bob's part after Alice performs the measurement with POVM element $e^{(1)}_{12}$ and communicates to Bob is given by
\begin{eqnarray}
\nonumber
\rho_3^{(1)} = \frac{1}{p_1} Tr_{12} [ e^{(1)}_{12} ( \ket{\psi} \bra{\psi} \otimes \ket{\Phi_n^+}\bra{\Phi_n^+}  ) e^{(1)}_{12} ],
\end{eqnarray}
where $p_1 = Tr_{123} [ e^{(1)}_{12} ( \ket{\psi} \bra{\psi} \otimes \ket{\Phi_n^+}\bra{\Phi_n^+}  ) e^{(1)}_{12} ] $.
Therefore, Bob's state can be expressed as
 \begin{eqnarray}
 \nonumber
  \rho_3^{(1)}&&=\frac{I}{2}+ 2i Im(\alpha \beta^*) \ket{0}\bra{1}- 2i Im(\alpha \beta^*) \ket{1}\bra{0}.
 \end{eqnarray}
\noindent This probability that the POVM $e^{(1)}_{12}$ will click is given by $p_{1}=\frac{2\lvert n \rvert^2}{(1+\lvert n \rvert^2)^2}$ .
  \\
  
\noindent The coherence of this state is given by
  
  \begin{eqnarray}
  \nonumber
    C_{l_1}(\rho^{(0)}_3)&&=2 \lvert Im(\alpha \beta ^*) \rvert\\
    \nonumber
   &&= 2\sin\phi\cos(\theta/2)\sin(\theta/2)\\
   \nonumber
 && \neq 2\cos(\theta/2)\sin(\theta/2)
 \end{eqnarray}
In this case if the partially known state happens to be from the polar circle of the Bloch sphere, then the coherence is perfectly teleported.\\
\noindent \textbf{Case II:}
Alice performs the following POVM to her part:
 \begin{eqnarray}
 \nonumber
  && \pi^{(0)}_{12}=\ket{\Phi_n^+}\bra{\Phi_n^+}+\ket{\Psi_n^+}\bra{\Psi_n^+},\\
  \nonumber
  && \pi^{(1)}_{12}=\ket{\Phi_n^-}\bra{\Phi_n^-}+\ket{\Psi_n^-}\bra{\Psi_n^-}.
 \end{eqnarray}
The state at Bob's part after Alice performs the measurement with POVM element $\pi^{(0)}_{12}$ and communicates to Bob is given by
\begin{eqnarray}
\nonumber
\rho_3^{(0)} = \frac{1}{p_0} Tr_{12} [ \pi^{(0)}_{12} ( \ket{\psi} \bra{\psi} \otimes \ket{\Phi_n^+}\bra{\Phi_n^+}  ) \pi^{(0)}_{12} ],
\end{eqnarray}
where $p_0 = Tr_{123} [ \pi^{(0)}_{12} ( \ket{\psi} \bra{\psi} \otimes \ket{\Phi_n^+}\bra{\Phi_n^+}  ) \pi^{(0)}_{12} ] $.
Therefore, Bob's state can be expressed as
 \begin{eqnarray}
  \nonumber
     &&\rho_3^{(0)}=\frac{1}{1+\lvert n \rvert^2} \ket{0}\bra{0}+\frac{\lvert n \rvert^2}{1+\lvert n \rvert^2} \ket{1}\bra{1}\\
     \nonumber
     && +\frac{\lvert n \rvert^2}{(1+\lvert n \rvert^2)(\lvert \alpha \rvert^2 +\lvert \beta \rvert^2 \lvert n \rvert^2 )} 2 Re(\alpha \beta^*) \ket{0}\bra{1}\\
     \nonumber
     &&+
     \frac{\lvert n \rvert^2}{(1+\lvert n \rvert^2)(\lvert \alpha \rvert^2 +\lvert \beta \rvert^2 \lvert n \rvert^2 )} 2 Re(\alpha \beta^*) \ket{1}\bra{0}.
 \end{eqnarray}
\noindent The probability that the POVM $\pi^{(0)}_{12}$ will click is given by $p_{0}=\frac{\lvert \alpha \rvert^2+ \lvert \beta \rvert^2 \lvert n \rvert^2}{(1+\lvert n \rvert^2)}$.\\
\noindent The coherence of the post-measurement state on Bob's side is given by 
\begin{eqnarray}
\nonumber
  C_{l_1}(\rho_3^{(0)})= \frac{4 \lvert n \rvert^2}{(1+\lvert n \rvert^2)(\lvert \alpha \rvert^2+ \lvert \beta \rvert^2 \lvert n \rvert^2)}   \lvert Re(\alpha \beta ^*) \rvert.
\end{eqnarray}
Here also, unlike the case discussed in the previous section, even if the partially known state is from the equatorial circle of the Bloch sphere, the coherence is not perfectly teleported. It is $\frac{4 \lvert n \rvert^2}{(1+\lvert n \rvert^2)(\lvert \alpha \rvert^2+ \lvert \beta \rvert^2 \lvert n \rvert^2)}   \lvert$ times the original coherence.\\

\noindent The state at Bob's part after Alice performs the measurement with POVM element $\pi^{(1)}_{12}$ and communicates to Bob is given by
\begin{eqnarray}
\nonumber
\rho_3^{(1)} = \frac{1}{p_1} Tr_{12} [ \pi^{(1)}_{12} ( \ket{\psi} \bra{\psi} \otimes \ket{\Phi_n^+}\bra{\Phi_n^+}  ) \pi^{(1)}_{12} ],
\end{eqnarray}
where $p_1 = Tr_{123} [ \pi^{(1)}_{12} ( \ket{\psi} \bra{\psi} \otimes \ket{\Phi_n^+}\bra{\Phi_n^+}  ) \pi^{(1)}_{12} ] $.
Therefore, Bob's state can be expressed as
  \begin{eqnarray}
  \nonumber
   && \rho_3^{(1)}=\frac{1}{1+\lvert n \rvert^2} \ket{0}\bra{0}+\frac{\lvert n \rvert^2}{1+\lvert n \rvert^2} \ket{1}\bra{1}\\
   \nonumber
     &&-\frac{\lvert n \rvert^2}{(1+\lvert n \rvert^2)(\lvert \alpha \rvert^2 \lvert n \rvert^2 +\lvert \beta \rvert^2)} 2 Re(\alpha \beta^*) \ket{0}\bra{1}\\
     \nonumber
     &&-
     \frac{\lvert n \rvert^2}{(1+\lvert n \rvert^2)(\lvert \alpha \rvert^2 \lvert n \rvert^2 +\lvert \beta \rvert^2)} 2 Re(\alpha \beta^*) \ket{1}\bra{0}.
  \end{eqnarray}
 \noindent  The probability that the POVM $\pi^{(1)}_{12}$ will click is $p_{1}=\frac{\lvert \alpha \rvert^2 \lvert n \rvert^2 + \lvert \beta \rvert^2 }{(1+\lvert n \rvert^2)}$. \\
 \\
  The coherence of this state is given by
  \begin{eqnarray}
  \nonumber
     C_{l_1}(\rho_3^{(1)})= \frac{4 \lvert n \rvert^2}{(1+\lvert n \rvert^2)(\lvert \alpha \rvert^2 \lvert n \rvert^2+ \lvert \beta \rvert^2 )}   \lvert Re(\alpha \beta ^*) \rvert.
  \end{eqnarray}
  Here, too, perfect teleportation of coherence is not possible, unlike in the case discussed in the previous section, even if the partially known state is from the equatorial circle of the Bloch sphere. It is $\frac{4 \lvert n \rvert^2}{(1+\lvert n \rvert^2)(\lvert \alpha \rvert^2 \lvert n \rvert^2+ \lvert \beta \rvert^2 )}   \lvert$ times the original coherence.\\
  \\
 \textbf{Case III:} In this case Alice performs the following POVM to her part:
 \begin{eqnarray}
 \nonumber
    && f^{(0)}_{12}=\ket{\Phi_n^+}\bra{\Phi_n^+}+\ket{\Phi_n^-}\bra{\Phi_n^-},\\
    \nonumber
    && f^{(1)}_{12}=\ket{\Psi_n^+}\bra{\Psi_n^+}+\ket{\Psi_n^-}\bra{\Psi_n^-}.
 \end{eqnarray}
  The state at Bob's part after Alice performs the measurement with POVM element $f^{(0)}_{12}$ and communicates to Bob is given by
\begin{eqnarray}
\nonumber
\rho_3^{(0)} = \frac{1}{p_0} Tr_{12} [ f^{(0)}_{12} ( \ket{\psi} \bra{\psi} \otimes \ket{\Phi_n^+}\bra{\Phi_n^+}  ) f^{(0)}_{12} ],
\end{eqnarray}
where $p_0 = Tr_{123} [ f^{(0)}_{12} ( \ket{\psi} \bra{\psi} \otimes \ket{\Phi_n^+}\bra{\Phi_n^+}  ) f^{(0)}_{12} ] $.
Therefore, Bob's state can be expressed as 
 \begin{eqnarray}
 \nonumber
    \rho_3^{(0)}&&=\frac{\lvert \alpha \rvert^2}{\lvert \alpha \rvert^2 + \lvert \beta \rvert^2 \lvert n \rvert^2} \ket{0}\bra{0}\\
    \nonumber
    && +\frac{\lvert \beta \rvert^2 \lvert n \rvert^2}{\lvert \alpha \rvert^2 + \lvert \beta \rvert^2 \lvert n \rvert^2} \ket{1}\bra{1}.
 \end{eqnarray}
 The probability that the POVM $f^{(0)}_{12}$ will click is given by $p_{0}=\frac{\lvert \alpha \rvert^2 + \lvert \beta \rvert^2 \lvert n \rvert^2}{1+\lvert n \rvert^2}$. \\
  In this case, the coherence of this state is zero. This is not useful for the teleportation of coherence.\\

 The state at Bob's part after Alice performs the measurement with POVM element $f^{(1)}_{12}$ and communicates to Bob is given by
\begin{eqnarray}
\nonumber
\rho_3^{(1)} = \frac{1}{p_1} Tr_{12} [ f^{(1)}_{12} ( \ket{\psi} \bra{\psi} \otimes \ket{\Phi_n^+}\bra{\Phi_n^+}  ) f^{(1)}_{12} ],
\end{eqnarray}
where $p_1 = Tr_{123} [ f^{(1)}_{12} ( \ket{\psi} \bra{\psi} \otimes \ket{\Phi_n^+}\bra{\Phi_n^+}  ) f^{(1)}_{12} ] $.
Therefore, Bob's state can be expressed as 
 \begin{eqnarray}
 \nonumber
  \rho_3^{(1)}=&&\frac{\lvert \beta \rvert^2}{\lvert \alpha \rvert^2 \lvert n \rvert^2 + \lvert \beta \rvert^2 } \ket{0}\bra{0}\\
  \nonumber
  && +\frac{\lvert \alpha \rvert^2 \lvert n \rvert^2}{\lvert \alpha \rvert^2 \lvert n \rvert^2 + \lvert \beta \rvert^2} \ket{1}\bra{1}.
 \end{eqnarray}
  The probability that the POVM $f^{(1)}_{12}$ will click is $p_{1}=\frac{\lvert \alpha \rvert^2 \lvert n \rvert^2 + \lvert \beta \rvert^2 }{1+\lvert n \rvert^2}$. \\
  
 \noindent We see that the coherence of the post-measurement state is zero.\\
So, we observe that only in the \textbf{Case I}, perfect teleportation of coherence is possible probabilistically, with the probability being $\frac{2\lvert n \rvert^2}{(1+\lvert n \rvert^2)^2}$ when the partially known states are from the polar circle of the Bloch sphere.
\section {Teleportation of Coherence with arbitrary Shared state and arbitrary POVM measurement:} 
In this section, we will talk about the teleportation of the coherence of an arbitrary state (pure and mixed) with the help of an arbitrary shared state and by considering in general arbitrary POVM measurements. Consider two finite-dimensional Hilbert space \(\cal H_1\) and \( \cal H_2\) over the field of complex numbers, with the dimension of $H_1$ being $n$. 
Let us consider a linear map $\Phi : \mathcal{B}(\cal H_1)\rightarrow \mathcal{B}(\cal H_2)$, where 
$B(\mathcal{H})$ is the vector space of bounded linear operators on $\mathcal{H}$.
Let $ \{e_{ij}=\ketbra{i}{j}\}^n_{i,j=1}$ be a complete set of matrix units for $B(\mathcal{H}_1)$. Then operator $\rho_\Phi =  \sum_{{i,j}=1}^{n} $ $e_{ij} \otimes \Phi(e_{ij})$ $\in $ $B(\cal H_1)\otimes B(\cal H_2) $ is known as the CJKS matrix \cite{book,article,CHOI1975285,kraus1983states,Sudarshan1985} 
corresponding to $\Phi$. It can easily be verified that the map $\Phi \rightarrow \rho_\Phi $ is linear and bijective and is called the CJKS isomorphism.\\ 
 
\noindent \textbf{CJKS theorem on completely positive maps~\cite{book,article,CHOI1975285,kraus1983states,Sudarshan1985}.
} \emph{The CJKS matrix $\rho_\Phi = $ $\sum_{{i,j}=1}^{n} $ $e_{ij} \otimes \Phi(e_{ij})$ $\in $ $B(\cal H_1)\otimes B(\cal H_2) $ is positive if and only if the map $\Phi : B(\cal H_1)\rightarrow B(\cal H_2)$ is completely positive.}\\
From the above theorem, we conclude that for any state $\tau$ and POVM $E$ belongs to $B(\cal H_1)\otimes B(\cal H_2) $ can be written as the following for two completely positive map $T$ and $\Phi_E$ 
\begin{eqnarray}
\tau &&= \sum_{{i,j}=1}^{n} e_{ij} \otimes T(e_{ij})\\
E &&=\sum_{{i,j}=1}^{n} e_{ij} \otimes \Phi_{E}(e_{ij})
\end{eqnarray}
In the following theorem, $\Phi^*_E$ denotes the completely positive map whose Kraus operators are the complex conjugate of that of the completely positive map $\Phi_E$.
\\
\\
\noindent \textbf{Theorem:} Let Alice have the unknown state $\rho$ whose coherence she needs to teleport to Bob. Let the shared state between Alice and Bob be $\tau$. If Alice performs a general measurement with
a POVM element E and communicates the result to Bob, then the state at Bob's lab is $\rho_B=T \circ \Phi^*_E(\rho)$, where $T$ is the completely positive map corresponding to the shared state $\tau$, $\Phi^*_E$ is the complex conjugate of the completely positive map corresponding to the POVM element $E$ and $``\circ"$ represents composition of the two maps.
\\
\proof The reduced state at Bob's location is $\rho_B=Tr_{12}((\sqrt{E}\otimes \mathbb{I})(\rho \otimes \tau) (\sqrt{E}\otimes \mathbb{I}))$. Let $A$ be any operator on the Hilbert space associated with Bob.
\\
\begin{widetext}
Now 
\begin{eqnarray}
Tr(\rho_B A)&&=Tr(Tr_{12}((\sqrt{E}\otimes \mathbb{I})(\rho \otimes \tau) (\sqrt{E}\otimes \mathbb{I}))A)  =Tr((\rho \otimes \tau)(E\otimes A)) \\
\nonumber
&&=\sum_{i,j,k,l} Tr[(\rho \otimes e_{ij} \otimes T(e_{ij})) (e_{kl} \otimes \Phi_E (e_{kl}) \otimes A)]\\
\nonumber
&&= \sum_{i,j,k,l} Tr[(\rho \otimes \ketbra{i}{j} \otimes T(\ketbra{i}{j}))(\ketbra{k}{l} \otimes \Phi_E (\ketbra{k}{l}) \otimes A)]\\
\nonumber
&&= \sum_{i,j,k,l} \sum_p \bra{l}\rho \ket{k} \bra{j}M_p \ket{k} \bra{l} M^{\dagger}_p \ket{i} Tr[T(\ketbra{i}{j}) A]\\
\nonumber
&&= \sum_{i,j,k,l} \sum_p \bra{i} M^{*}_p \ket{l} \bra{l}\rho \ket{k} \bra{k}M^{t}_p \ket{j} Tr[T(\ketbra{i}{j}) A]\\
\nonumber 
&&= \sum_{i,j} \bra{i} \sum_{p} M^{*}_p \rho (M^{*}_p)^{\dagger} \ket{j} Tr[T(\ketbra{i}{j}) A]\\
\nonumber
&&= \sum_{i,j} \bra{i} \Phi^{*}_{E}(\rho)\ket{j} Tr[T(\ketbra{i}{j}) A]\\
\nonumber
&&= Tr[T(\Phi^{*}_{E}(\rho))A]
\end{eqnarray}
\end{widetext}
As $A$ is arbitrary, we conclude that
\begin{eqnarray}
    \rho_B =T \circ \Phi^*_E(\rho)
\end{eqnarray}
\subsection{Teleportation of coherence using mixed entangled state as a resource :}
In this subsection, we study the teleportation of coherence using mixed entangled states as a resource state, unlike pure entangled states in the previous subsections. Here, we consider two types of mixed entangled states as examples of resources that can be used. These are (a) Maximally mixed entangled states (b) Werner states.\\ 

\noindent \textbf{Maximally Entangled Mixed States:} This class of two qubit states appeared first in \cite{Maximally,Maximally1}. It has the maximum amount of entanglement because no global unitary transformation can increase the entanglement of formation or even the negativity of entanglement of the states in this class. Let Alice and Bob share a two-qubit maximally entangled mixed state given by the following:
\begin{eqnarray}
\tau= p_1 \ketbra{\Psi^-}{\Psi^-}+p_2 \ketbra{00}{00}+p_3 \ketbra{\Psi^+}{\Psi^+}+p_4 \ketbra{11}{11} \nonumber
\end{eqnarray}
where $p_1, p_2, p_3$ and $p_4$ are probabilities in decreasing order, i.e., $p_1 \geq p_2\geq p_3\geq p_4$.
Alice performs the following POVMs on the input and half of the entangled pair 
 \begin{eqnarray}
  \Pi^{(0)}_{12} =\ket{\Phi^+}\bra{\Phi^+} + \ket{\Psi^+}\bra{\Psi^+}, \label{POVM 0}
 \end{eqnarray}
 \begin{eqnarray}
    \Pi^{(1)}_{12}=\ket{\Phi^-}\bra{\Phi^-}+\ket{\Psi^-}\bra{\Psi^-}. \label{POVM 1}
 \end{eqnarray} 
Let Alice perform the measurement given by the POVM element $\Pi^{(0)}_{12}$ she communicate her result to Bob. In this case, $\Phi^*_E$ and $T$ are given by the following equations:
\begin{eqnarray}
\nonumber
   \Phi^*_E(A) &&=A+\sigma_x A \sigma_x \\
   \nonumber
   T(A) &&=p_1 \sigma_y A \sigma_y +p_2 \bra{0}A\ket{0} \ketbra{0}{0}\\
   \nonumber
   && +p_3 \sigma_x A \sigma_x +p_4 \bra{1}A\ket{1} \ketbra{1}{1} \nonumber
\end{eqnarray}
where $A$ belongs to the set of bounded linear operators on the Hilbert space associated with Bob.
The state of the particle 
at Bob's lab is given by
\begin{eqnarray}
    \nonumber
   \rho_B=T\Phi^*_E(\rho)=\frac{1}{1+p_1+p_3} (p_1 \sigma_z \rho \sigma_z +p_1 \sigma_y \rho \sigma_y \\
   + p_3 \sigma_x \rho \sigma_x + p_3 \rho + p_2 \ketbra{0}{0}+p_4 \ketbra{1}{1} ). \label{nice}
\end{eqnarray}
The coherence of this state is given by
\begin{eqnarray}
   C(\rho_B)=\frac{4 \modu{p_1 -p_3}}{1+p_1+p_3} \modu{Re(\rho_{01})}.
\end{eqnarray}
\\
The concurrence of the above-mentioned maximally entangled mixed state is 
\begin{eqnarray}
   \mathcal{C}(\tau)= max\{0, p_1-p_3 -\sqrt{p_2 p_4}\}
\end{eqnarray}
Assuming $p_4=0$ the concurrence reduces to $\mathcal{C}(\tau)=p_1-p_3$.
 From Eq. (\ref{nice}) we get 
 \begin{eqnarray}
 \nonumber
    C(\rho_B) &&= \frac{2\mathcal{C}(\tau)}{1+p_1 +p_3} 2\modu{Re(\rho_{01})}\\
    \nonumber
    && \leq \frac{2\mathcal{C}(\tau)}{1+p_1 +p_3} C(\rho)\\
    && \leq \frac{2\mathcal{C}(\tau)}{1+\mathcal{C}(\tau)} C(\rho) \label{bigboss}
 \end{eqnarray}
 \\
 When POVM $\Pi_{12}^{(1)}$ clicks and Alice communicates the result to Bob, then the state at Bob's place is the same as the Eq. (\ref{nice}) and hence the relation between the coherence at Bob's place and the unknown state obeys the Eq. (\ref{bigboss})\\
 
\noindent \textbf{Wener States:} Now let's assume that Alice and Bob share the Warner state $\tau= p \ketbra{\Psi^-}{\Psi^-}+\frac{1-p}{4} \mathbb{I}$ with $0 \leq p \leq 1$ and Alice performs a measurement using the POVM described by Eq.(\ref{POVM 0}) and Eq. (\ref{POVM 1}). If POVM given by Eq. (\ref{POVM 0}) clicks and Alice communicates the result to Bob, the state at Bob's place is given by
 \begin{eqnarray}
    \rho_B=p\sigma_y \rho \sigma_y + p \sigma_z \rho \sigma_z + \frac{1-p}{2} \mathbb{I}.
 \end{eqnarray}
 The coherence of the above state with respect to the computational basis is as follows:
 \begin{eqnarray}
    C(\rho_B)=\frac{2p}{1+p} \modu{2 Re(\rho_{01})} .\label{W coherence}
 \end{eqnarray}
If the POVM described by Eq. (\ref{POVM 1}) clicks and Alice communicates the result to Bob, then the state at Bob's place is given by
 \begin{eqnarray}
     \rho_B=p\sigma_x \rho \sigma_x + p \rho + \frac{1-p}{2} \mathbb{I}.
 \end{eqnarray}
The Coherence of the above state is the same as in Eq. (\ref{W coherence}).\\
It is a well-known fact that the Warner state is separable when $p \leq \frac{1}{3}$.
From the Eq. (\ref{W coherence}) we clearly see that the Teleported coherence is non zero for $0 <p \leq \frac{1}{3}$. Hence, we conclude that it is possible to teleport coherence without entanglement.

 \subsection{Teleportation of coherence of an unknown mixed state:} 
 
 At last, we consider the case of teleportation of coherence of an unknown mixed state with maximally entangled state as a resource and POVM described by Eq. (\ref{povmI}) and (\ref{povmII}). In this case the completely positive map $T$ is the identity map and the map $\Phi^*_{\Pi^{(0)}_{12}}$ and $\Phi^*_{\Pi^{(1)}_{12}}$ is given by the following equations,
 \begin{eqnarray}
     \Phi^*_{{\Pi^{(0)}_{12}}}(\rho) &&=\rho+ \sigma _{x} \rho  \sigma _{x}\\
     \Phi^*_{{\Pi^{(1)}_{12}}}(\rho) &&=\sigma _z \rho \sigma_z + \sigma _{y} \rho  \sigma _{y}
 \end{eqnarray}
 So the state at Bob's place when the outcomes of the measurements of the POVMs ${\Pi^{(0)}_{12}}$ and ${\Pi^{(1)}_{12}}$ are communicated to Bob is given by:
 \begin{eqnarray}
     \rho^{(0)}_B=\Phi^*_{{\Pi^{(0)}_{12}}}(\rho) &&=\rho+ \sigma _{x} \rho  \sigma _{x},\\
    \rho^{(1)}_B= \Phi^*_{{\Pi^{(1)}_{12}}}(\rho) &&=\sigma _z \rho \sigma_z + \sigma _{y} \rho  \sigma _{y}.
 \end{eqnarray}
 In both cases, the coherence w.r.t. the computational basis turns out to be equal and is given by:
 \begin{eqnarray}
     C( \rho^{(0)}_B)=C( \rho^{(1)}_B)&&= \modu{\rho_{01}+\rho_{10}}\\
     && \leq \modu{\rho_{01}}+ \modu{\rho_{10}}\\
     && \leq C(\rho)
 \end{eqnarray}
 Clearly, the teleported coherence is less than or equal to the original coherence, even in the more general scenario when the unknown state is a mixed state. However, if we have partial knowledge about the state, i.e., the matrix elements of the state are real, then the teleported coherence is equal to the original coherence.\\
 
\section{Conclusion}

Quantum Coherence being a useful resource in various information processing tasks, it is of utmost importance to transfer the coherence of a coherent state to a state where there is no coherence. This is required in a quantum network where one quantum processor can teleport the coherence to another quantum processor if there is a requirement. Therefore, it is important to know if it is possible to teleport coherence without actually transferring the entire information of the state. Interestingly, we are able to show that this can be done with a lesser number of cbits for the states that are taken from equatorial and polar circles.

In summary, we have addressed the question of whether we can teleport the coherence of an unknown state with the transfer of a lesser number of classical bits. We show that by transferring one cbit of information, we can teleport the coherence only for states which lie on the equatorial and polar circles. This is done when we have a maximally entangled state as a resource. However, for a non-maximally entangled state, we can teleport coherence, if not deterministically but with a certain probability of success. We have also investigated the possibility of teleportation of coherence using two qubit maximally entangled mixed states and Werner states as resources. We observe that even when the Werner state is separable, the teleported coherence is non-zero, which indicates that the teleportation of coherence is possible without entanglement. In a nutshell, we are able to provide a cost-effective way to teleport coherence as a resource to a place where there is a requirement of this resource. This is done without actually transferring the state. 

\noindent \textit{Acknowledgement:} Authors acknowledge  Dr. S. Mitra and Dr. K. Srinathan for having various useful discussions. S.~Patro is currently affiliated to QuSoft, CWI and University of Amsterdam, and is supported by the Robert Bosch Stiftung, however, this work was done while at International Institute of Information Technology-Hyderabad, India. V. Aradhya is currently affiliated to National University of Singapore and he contributed to this project while he was an undergraduate student at International Institute of Information Technology-Hyderabad, India.

\end{document}